\documentclass[twocolumn,showpacs,aps,prl,superscriptaddress]{revtex4}

\usepackage{graphicx}
\usepackage{dcolumn}
\usepackage{amsmath}
\usepackage{epsfig}
\usepackage{color}

\input babarsym
\input colordvi.tex

\def\ORL{79.9}
\def\NBB{\ensuremath{87.8\times 10^6}\xspace}
\def\Nseen{13\null\xspace}
\def\TheLimitAll{2.5}
\def\TheLimit{\ensuremath{\TheLimitAll\times10^{-5}}\xspace}
\def\TheLimitOne{\ensuremath{ 3.4}\xspace}
\def\TheLimitTwo{\ensuremath{ 3.5}\xspace}
\def\TheLimitThr{\ensuremath{ 3.9}\xspace}
\def\TheLimitFou{\ensuremath{ 8.0}\xspace}
\def\TheLimitFiv{\ensuremath{14.8}\xspace}
\def\TheLimitSix{\ensuremath{20.3}\xspace}
\def\TheCVAll{\ensuremath{1.0^{+1.1}_{-0.9}}}
\def\TheCV{   \ensuremath{(\TheCVAll)\times10^{-5}}\xspace}
\def\TheSignificance{0.86}

\def\geantfour{\mbox{\tt GEANT4}\xspace}

\def\kpipiz{\ensuremath{\Km\pip\piz}\xspace}
\def\kpipipi{\ensuremath{\Km\pip\pip\pim}\xspace}

\def\Dstarzpiz{\ensuremath{\Dstarz\piz}\xspace}

\def\mydecay{\ensuremath{\Bzb\to\Dstarz\g}\xspace}
\def\pizbackground{\ensuremath{\Bzb\to\Dstarz\piz}\xspace}
\def\etal{\hbox{\em et al.\null}}
\def\BaBar{\babar\xspace}
\def\p{\phantom{0}}
\newcommand{\BABARPubYear}     {05}
\newcommand{\BABARPubNumber}   {009}
\newcommand{\SLACPubNumber} {11292}
\newcommand{\LANLNumber}  {0506070}

\begin{document}

\begin{flushleft}
\BaBar-PUB-\BABARPubYear/\BABARPubNumber \\
SLAC-PUB-\SLACPubNumber\\
hep-ex/\LANLNumber\\
DOI: 10.1103/PhysRevD.72.051106 \\
Published as \jprd{72}, 051106 (2005) \\ [12mm]
\end{flushleft}

\title{
{\large \bf \vspace{-4mm} 
Search for the rare decay \mydecay}
}

%% author list as of 02-Feb-2005 (631 authors)
%
\author{B.~Aubert}
\author{R.~Barate}
\author{D.~Boutigny}
\author{F.~Couderc}
\author{Y.~Karyotakis}
\author{J.~P.~Lees}
\author{V.~Poireau}
\author{V.~Tisserand}
\author{A.~Zghiche}
\affiliation{Laboratoire de Physique des Particules, F-74941 Annecy-le-Vieux, France }
\author{E.~Grauges}
\affiliation{IFAE, Universitat Autonoma de Barcelona, E-08193 Bellaterra, Barcelona, Spain }
% ...
\collaboration{\babar\ Collaboration}
\noaffiliation

\date{\today \\ aaPhys.~Rev.~D: Received 27 June 2005; published 29
September 2005}

\begin{abstract}
We report on a search for the rare decay $\kern 0.18em\overline{\kern
  -0.18em B}{}^0\rightarrow D^{*0}\gamma$, which in the standard model
is dominated by $W$-exchange.  The~analysis is based on a data sample
comprising $87.8\times 10^6$ $B\kern 0.18em\overline{\kern -0.18em
  B}$\xspace pairs collected with the \mbox{\slshape
  B\kern-0.1em{\smaller A}\kern-0.1em B\kern-0.1em{\smaller A\kern-0.2em
    R}}\xspace detector at the \hbox{PEP-II} asymmetric-energy $B$
Factory at~SLAC\null.  No~significant signal is observed, and an upper
limit on the branching fraction of $2.5\times10^{-5}$\xspace at the 90\%
confidence level is obtained.
\end{abstract}


\maketitle

\begin{figure}[bht]
\begin{center}
  \centerline{\epsfxsize 2.5 truein \epsfbox{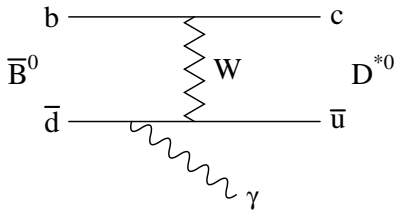}}
\end{center}
\vspace{-1.1cm}
\caption[]{\label{fig:myDecay}\small \W-exchange is the leading
contribution to the \mydecay decay in the standard model.  The~photon
may be emitted from any quark line or the~\W$\!\!\!$.}
\end{figure}

Within the standard model~(SM), the rare decay \mydecay~\cite{bib:CC} is
dominated by the \W-boson exchange process.  One of the leading SM
contributions to the decay is illustrated in Fig.~\ref{fig:myDecay}.
Similar \W-exchange transitions are present in other decays.  For
example, they contribute to the decay $\Bz\to\rho^0\g$ along with the
leading electromagnetic-penguin process~\cite{bib:HYChen}.  The
branching fraction $\BR(\mydecay)$ is estimated to be of order
$10^{-6}$~\hbox{\cite{bib:HYChen,bib:HYChenEtAl,bib:RRMendel}}, but the
presence of a large $\qqbar g$ (color octet) component in the wave
function of the \B meson may reduce the color-suppression enough to
raise the branching fraction by a factor of about
10~\cite{bib:RRMendel}.  A~search for \mydecay, published by the CLEO
collaboration~\cite{bib:CLEOPRL84-4292}, resulted in a limit of
$\BR(\mydecay)<5.0\times10^{-5}$ at the 90\% confidence level~(C.L.).

We search for the decay \mydecay in data collected using the \BaBar
detector operating at the Stanford Linear Accelerator Center (SLAC)
PEP-II asymmetric-energy \epem collider.  The collider runs with
a~center-of-mass (CM) energy of 10.58\gev at the peak of the \FourS
resonance, which decays into \BpBm and \BzBzb pairs.  The analysis is
based on \NBB \BB pairs, corresponding to an integrated luminosity of
\ORL\invfb.  The \BaBar detector is described in detail in
Ref.~\cite{bib:NIM}; here we introduce briefly the detector systems
important for the present analysis.  Tracks of charged particles and
their momenta are measured in a vertex tracker, consisting of five
layers of double-sided silicon microstrip detectors, and a 40-layer
drift chamber.  Both systems are located within a 1.5-T solenoidal
magnetic field and provide d$E$/d$x$ measurements for particle
identification~(PID).  A~Cherenkov ring imaging detector adds
measurements for PID by recording Cherenkov light emitted from charged
particles traversing transparent quartz bars.  Photons~are identified by
an~electromagnetic calorimeter consisting of 6580~CsI(Tl) crystals.

Event samples from Monte Carlo (MC) simulations are used to optimize the
event selection criteria and to estimate the signal efficiency and
background.  The~detector response is simulated using
\geantfour~\cite{bib:GEANT4}.  The MC sample for the signal \mydecay
contains 328$\,$000 events.  We~use MC samples of similar size for
several exclusive \B-decay background modes.  The color-suppressed
hadronic decay \pizbackground, with branching fraction
$(2.7\pm0.5)\times10^{-4}$~\cite{bib:PDG}, is the largest contributor
among them.  Other backgrounds originate from \BB modes with
incompletely or incorrectly reconstructed particles, and from random
combinations of particles from two different \B mesons or from \qqbar
pairs.  For these, we use MC samples of generic \BB events and continuum
\qqbar ($q =u,d,s,c$) events corresponding to about 200\invfb and
110\invfb, respectively.

The \Dstarz candidates are reconstructed in six submodes, with
$\Dstarz\to\Dz(\piz,\g)$ and
$\Dz\to(\Km\pip,\Km\pip\piz,\Km\pip\pip\pim)$.  The~event selection
criteria are optimized by using the MC samples to maximize $S^2/(S+B)$,
where $S$ ($B$) is the number of signal (background) events.  A~signal
branching fraction of $10^{-6}$ is assumed during the optimization.
The~most important selection requirements are described below.

The photon from the decay \mydecay is emitted with an energy of about
2.3\gev in the CM frame (``hard photon'').  Although this high energy
leads to a relatively clear signal, care must be taken that remnants of
\piz decays are not mistaken as the signal photon.  The ``\piz veto''
rejects a hard photon candidate if its combination with any other photon
with laboratory energy larger than 30\mev yields an invariant mass in
\hbox{the range $[110,155]\mevcc$.}  A~similar veto for $\eta$ decays
rejects a photon candidate if its combination with any other photon of
laboratory energy larger than 250\mev yields an invariant mass within
$[508,588]\mevcc$.  Hard photon candidates must also pass a calorimeter
shower-shape requirement designed to exclude irregularly shaped showers
caused, for example, by overlapping photons from \piz~decay.  Background
is further suppressed by requiring a hard photon candidate to be
isolated from all other showers and tracks by at least 50\cm in the
calorimeter.

A~photon candidate from the decay $\Dstarz\to\Dz\g$ (``soft photon'')
must satisfy the same shower-shape requirement and $\eta$~veto that are
applied to hard photons.  In~the \piz veto the minimum energy for the
other photon is raised to 80\mev and the invariant mass range is
restricted to $[115,150]\mevcc$.  In addition, the CM energy of the soft
photon candidate has to be at least~$110\mev$.

The~mass of the \piz in the decay $\Dstarz\to\Dz\piz$ and of the \piz in
the decay $\Dz\to\kpipiz$ is required to be within 11\mevcc of the true
\piz mass (which corresponds to a cut at about 1.7$\,\sigma$, where
$\sigma$ is the \piz mass resolution).  Photons from \piz decay need a
minimum energy of 30\mev and have to pass a similar, but slightly less
stringent, shower-shape requirement as the hard and soft photons.

The charged \kaon and $\pi$ tracks are required to originate from the
interaction point and have to pass likelihood-based particle
identification selections using d$E$/d$x$ and Cherenkov light
measurements.  The \kaon track in the $\Dz\to\kpipipi$ decay is in
addition required to have a transverse momentum larger than 0.1\gevc and
at least 12~hits in the drift chamber.  A~vertex fit is applied to the
\Dz candidates.  They are required to have masses close to the known \Dz
mass: within 12\mevcc ($\sim 1.8\,\sigma$) for $\Dz\to\Km\pip$, within
23\mevcc ($\sim1.9\,\sigma$) for $\Dz\to\Km\pip\piz$, and within
12\mevcc ($\sim2.3\,\sigma$) for $\Dz\to\Km\pip\pip\pim$.  Additional
selection requirements are applied to $\Dz$ candidates decaying into
$\Km\pip\piz$.  The~laboratory energy of the \piz must be at least
250\mev, and only $\Dz\to\kpipiz$ candidates that appear in the Dalitz
plot close to known resonances~\cite{bib:dalitz} are accepted.  The
difference between the \Dstarz and \Dz mass has to be within 2\mevcc
($\sim2\,\sigma$) for $\Dstarz\to\Dz\piz$ and within 9\mevcc
($\sim1.8\,\sigma$) for $\Dstarz\to\Dz\g$ of the known value of
Ref.~\cite{bib:PDG}.

The \Dstarz helicity angle $\theta^*_H$ is defined in the \Dstarz CM
frame as the angle between the direction of the \Dz and the direction
opposite to the \B momentum.  For the $\Dstarz\to\Dz\piz$ modes,
$\cos\theta^*_H$ is distributed as $\sin^2\theta^*_H$ for signal, but as
$\cos^2\theta^*_H$ for background from \pizbackground.  Optimization
leads to the requirement $|\cos\theta^*_H|<0.75$.  No~such condition is
imposed for $\Dstarz\to\Dz\g$ modes.

Several selection requirements reduce the number of fake decays from
\qqbar continuum background.  The angle $\theta^*_B$ is defined as the
angle between the \B candidate momentum in the \FourS CM frame and the
beam axis.  In~\qqbar background events the distribution is uniform in
$\cos\theta^*_B$, while for real \B mesons it follows a
$\sin^2\theta^*_B$ distribution.  We require that
$|\cos\theta^*_B|<0.8$.  The angle $\theta^*_T$ is the angle between the
thrust direction of the \B candidate and the thrust direction computed
from the other photons and tracks in the event.  For signal events the
distribution of $|\cos\theta^*_T|$ is flat, while for continuum events
the distribution has a maximum at~$|\cos\theta^*_T| = 1$ due to their
jetlike nature.  We~require that $|\cos\theta^*_T|<0.75$.

The~candidates are subsequently characterized with two kinematic
quantities, \mes and~\DeltaE.  For~the ``energy-substituted mass'' \mes,
the energy of the \B candidate is substituted by precisely known beam
parameters:
\begin{eqnarray}
    \mes=\sqrt{\left(s/2+c^2{\bf p}_0\cdot{\bf p}_B\right)^2/E^2_0-c^2{\bf p}^2_B}\,,
\end{eqnarray}
where $s$ is the square of the total CM energy, $E_0$ and ${\bf p}_0$
are the energy and momentum of the initial \FourS in the laboratory
frame, and ${\bf p}_B = {\bf p}_{\Dstarz} + {\bf p}_{\g}$ is the
momentum of the \B candidate, also taken in the laboratory frame.  The
quantity \DeltaE is defined as the difference between the energy of the
\B candidate $E^*$ and the beam energy, both taken in the CM system:
\begin{eqnarray}
    \DeltaE=E^*-\frac{1}{2}\sqrt{s}\,.
\end{eqnarray}
Requirements of $|\DeltaE|<0.34\gev$ and $5.2<\mes<5.29\gevcc$ are
applied at this point.

If an event contains more than one \mydecay candidate passing all
selection criteria, the selection is made based on a $\chi^2$ function
that uses the measured \Dz mass and \hbox{\Dstarz-\Dz} mass difference,
the measured resolutions, and known mass and mass-difference values from
Ref.~\cite{bib:PDG}.  This~selection is sufficient, as the ambiguity is
never due to the presence of two hard photon candidates.

\begin{figure}[hbt]
\centerline{\epsfxsize3.3in \epsfbox{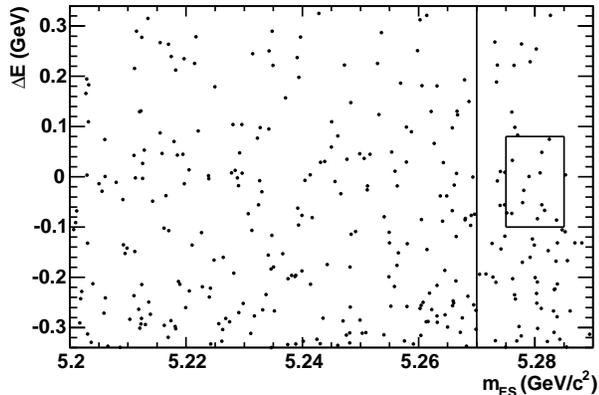}}
\vskip -.6 cm
\caption[]{\label{fig:DeltaE-mES_onRes}\small Distribution of data
events in the \DeltaE-\mes plane.  The~lines indicate the regions of the
signal box and of the grand sideband.}
\end{figure}

The distribution of \mes versus \DeltaE is shown in
Fig.~\ref{fig:DeltaE-mES_onRes} for the data taken at the \FourS
resonance.  While the combinatorial \qqbar background is smoothly
distributed over this plane, the signal should peak around $\DeltaE=0$
and $\mes=5.28\gevcc$.  The borders of the signal box are given by
$5.275<\mes<5.285\gevcc$ and $-0.1<\DeltaE<0.08\gev$, extending to about
1.7 (1.9) times the resolution of \mes (\DeltaE) of signal events.
The~\DeltaE constraint is asymmetric to account for the energy leakage
from the calorimeter for the hard photon candidates.  The area with \mes
ranging from $5.2\gevcc$ to $5.27\gevcc$ is called the ``grand
sideband.''

The~contributions to the systematic uncertainties in the signal
reconstruction efficiencies are listed in Table~\ref{tab:systerr}.
The~overall relative uncertainties range from 16.5\% to~19.8\%,
depending on the reconstruction mode~(see Table~\ref{tab:results}).  The
major contributors are described here in more detail. The uncertainties
in the photon reconstruction due to efficiency, energy scale, and energy
resolution uncertainties are studied with control samples and result in
an uncertainty of 2.5\% per photon (5\% per \piz).  Studies of the track
finding efficiency using control samples result in uncertainties of
2.6\% to 5.9\% depending on the mode.  The~size of the uncertainty in
the \DeltaE and \mes selection is obtained by varying the selection
according to observed differences between data and MC simulation.  For
the thrust angle~$\theta^*_T$, the \Bz angle~$\theta^*_B$, and the
heli\-city angle~$\theta^*_H$, the size of the uncertainties is obtained
by shifting the selection requirement by $\pm0.05$ in the cosine of each
angle.  The~uncertainty due to possible discrepancies between data and
MC simulation in the \Dz mass and the \Dstarz-\Dz mass difference is
estimated by comparing these distributions for events in the grand
sideband.  Data and Monte Carlo simulation agree sufficiently well, and
the size of the systematic uncertainty in the efficiency is obtained
from the uncertainty on the fits to the mass and mass-difference plots.

Several correction factors are applied to the signal efficiency based on
comparison studies on data and Monte Carlo simulations: a~tracking
efficiency factor of 0.992 for the kaon in the decay $\Dz\to\kpipipi$, a
factor 0.95 for the decay $\Dz\to\kpipiz$ due to the selection
requirement involving the Dalitz structure, and factors from 0.89 to
0.95 depending on the reconstructed mode due to photon reconstruction.
The overall selection efficiencies for the six signal modes are listed
in Table~\ref{tab:results}.  The~uncertainties on the efficiencies
include all contributions from systematic effects on the efficiencies.
The combined efficiency (weighted by the branching fractions of the
individual modes and taking correlations in the uncertainties between
the six submodes into account) is~$(1.8\pm0.3)\%$.  In~the determination
of the \mydecay branching fraction results, a 1.1\% uncertainty on the
number of \BB pairs in the data sample is included as well as the
contribution by the \Dz (\Dstarz) branching fraction
uncertainties~\cite{bib:PDG}.

\begin{table}[htb]
 \caption{Maximal and minimal relative systematic uncertainties in the
efficiency for the individual reconstruction modes.\label{tab:systerr}}
\begin{tabular*}{\hsize}{@{}l@{\extracolsep\fill}c@{}}
\hline\hline     
                            & Systematic uncertainty        \\ 
                            & in \% of the efficiency       \\ \hline
\g \& \piz reconstruction   & 5.0 to 12.5 \rule{0pt}{10pt}  \\
Hard \g separation          & 2.0                           \\
Shower shape                & 1.0 to 2.5                    \\
\piz, $\eta$ veto           & 1.5 to 3.0                    \\
Track finding efficiency    & 2.6 to 5.9                    \\
Kaon PID                    & 3.0                           \\
\Dz mass                    & 2.3 to 4.4                    \\   
\Dstarz-\Dz mass difference & 2.5 to 6.7                    \\          
Dalitz structure            & 0.0 to 5.0                    \\
Helicity angle $\theta^*_H$ & 0.0 to 3.8                    \\
Thrust angle $\theta^*_T$   & 5.5 to 7.3                    \\
\Bz angle $\theta^*_B$      & 3.0 to 3.8                    \\
\DeltaE                     & 8.6 to 12.0                   \\
\mes                        & 2.3 to 4.0                    \\
Simulation statistics       & 2.0 to 4.7                    \\ [0.75ex]
Sum                         &16.5 to 19.8 \rule{0pt}{9.5pt} \\
\hline \hline
\end{tabular*} \end{table}

The~number of events expected in the signal box due to background is not
estimated from data, but from MC simulation, since the \DeltaE-\mes
distributions of several categories of \BB background peak inside the
signal box.  After~counting the MC events and scaling the number to
\ORL\invfb, a total of $9.4\pm1.7$ background events is expected for all
six modes combined.  Of~those, 2.9 events originate from \pizbackground,
5.1~events from other \BB decays, and 1.4~events from \qqbar~events.
The breakdown for each channel is given in Table~\ref{tab:results}.

The~estimate of the number of background events is cross-checked by two
studies, one based on events in the grand sideband, and the other based
on events in the signal box using a control sample of \Dstarzpiz events.
The first study results in ratios of data-to-MC events ranging from
$1.0\pm0.3$ to $1.5\pm0.2$ for the various \Dstarz decay modes, and a
ratio of $1.2\pm0.1$ for all modes combined.  Taking the uncertainties
into account, data and MC simulation do not disagree significantly.
For~the second study, \pizbackground events are selected by loosening
some selection requirements and by inverting the \piz veto: we now keep
events in which a photon combined with the hard photon forms a
reasonable \piz candidate.  The~number of events seen in the signal box
is usually found to be lower in data than in MC simulation with
data-to-MC ratios from $0.3\pm0.3$ to $1.2\pm0.7$ for the various
\Dstarz decay modes and $0.6\pm0.2$ for all modes combined.

\begin{table*}[thb]
 \caption{Results for individual modes and all modes combined.  The
upper limit is given for 90\% C.L.\label{tab:results}}
\begin{tabular*}{\hsize}{@{}l@{\extracolsep\fill}@{\hskip-2mm}ccccccc@{}}
\hline\hline
                                   & Branching fraction     & Relative systematic &  Signal     & Expected    & Range of   & Observed in & Branching fraction \\
                                   & of mode \cite{bib:PDG} & uncertainty    & efficiency  & background  & data-to-MC & signal box  & upper limit        \\
Mode                               & (in \%)                & (in \%)        & (in \%)     & (events)    &  ratios    & (events)    & ($\times 10^{-5}$) \\
\hline 
$\Dstarz\to\Dz\piz$                & \rule{0pt}{10pt}       &                &             &             &            &             &                    \\ 
$\Dz\to\Km\pip$                    & \p2.3                  & 16.5           & $4.2\pm0.7$ & $1.5\pm0.7$ & 0.0 to 1.6 &  \p1        & \p\TheLimitOne     \\
$\Dz\to\Km\pip\piz$                & \p7.9                  & 19.8           & $1.2\pm0.2$ & $2.0\pm0.8$ & 0.0 to 1.3 &  \p1        & \p\TheLimitTwo     \\
$\Dz\to\Km\pip\pip\pim$            & \p4.6                  & 17.3           & $2.0\pm0.3$ & $0.7\pm0.1$ & 0.5 to 2.0 &  \p1        & \p\TheLimitThr     \\ 
\hline 
$\Dstarz\to\Dz\g$                  & \rule{0pt}{10pt}       &                &             &             &            &             &                    \\ 
$\Dz\to\Km\pip$                    & \p1.4                  & 17.3           & $3.8\pm0.7$ & $1.6\pm0.4$ & 0.4 to 1.6 &  \p2        & \p\TheLimitFou     \\
$\Dz\to\Km\pip\piz$                & \p4.9                  & 19.6           & $0.9\pm0.2$ & $2.4\pm1.2$ & 0.1 to 1.2 &  \p3        &   \TheLimitFiv     \\
$\Dz\to\Km\pip\pip\pim$            & \p2.8                  & 17.7           & $1.7\pm0.3$ & $1.2\pm0.2$ & 0.2 to 1.7 &  \p5        &   \TheLimitSix     \\ 
\hline 
All modes combined\rule{0pt}{10pt} &  23.9                  & 16.8           & $1.8\pm0.3$ & $9.4\pm1.7$ & 0.4 to 1.3 &   13        & \p\TheLimitAll     \\
\hline\hline
\end{tabular*} \end{table*}

\begin{figure}[hbt]
\centerline{\epsfxsize 1.7in \epsfbox{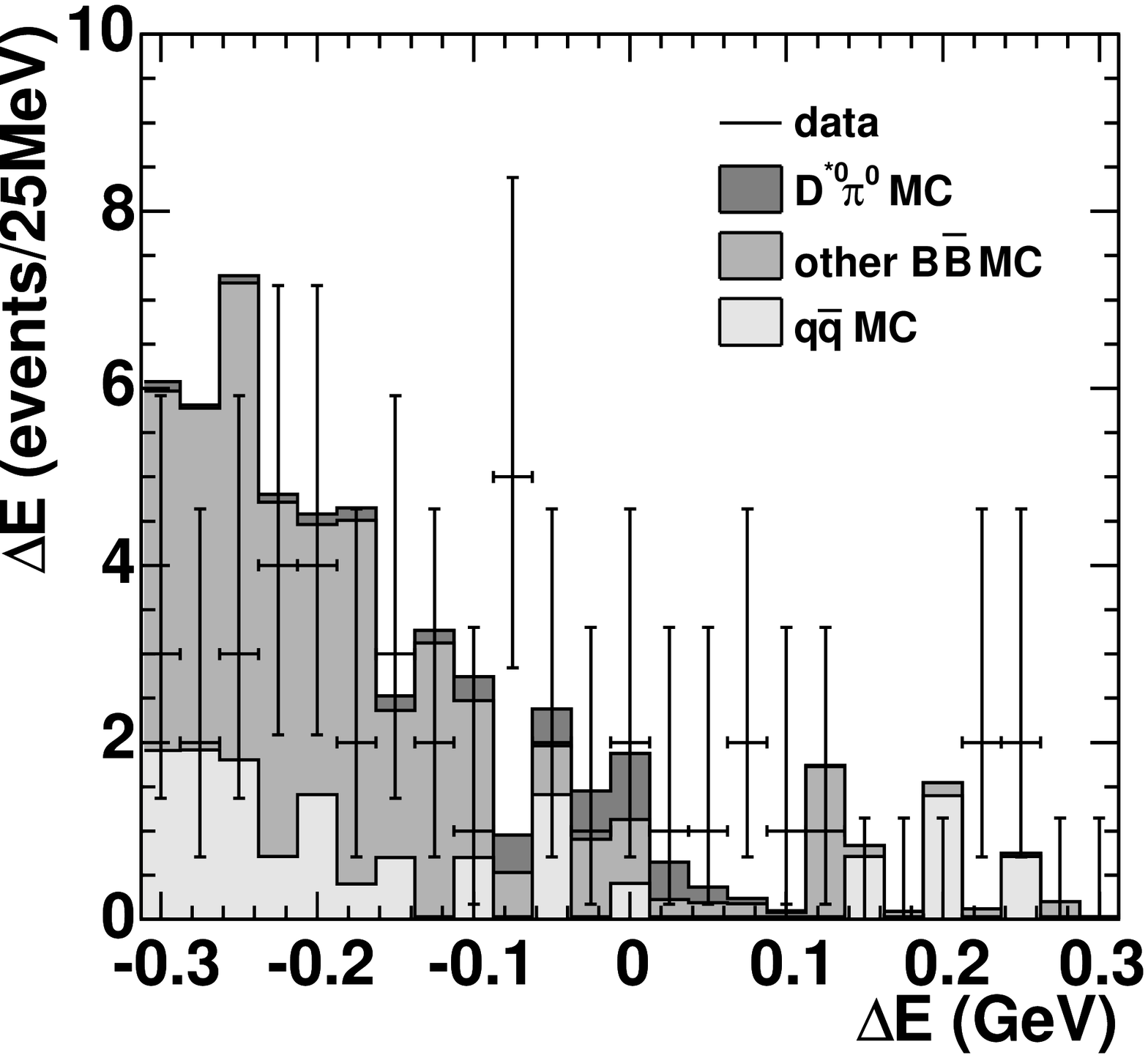}
            \epsfxsize 1.7in \epsfbox{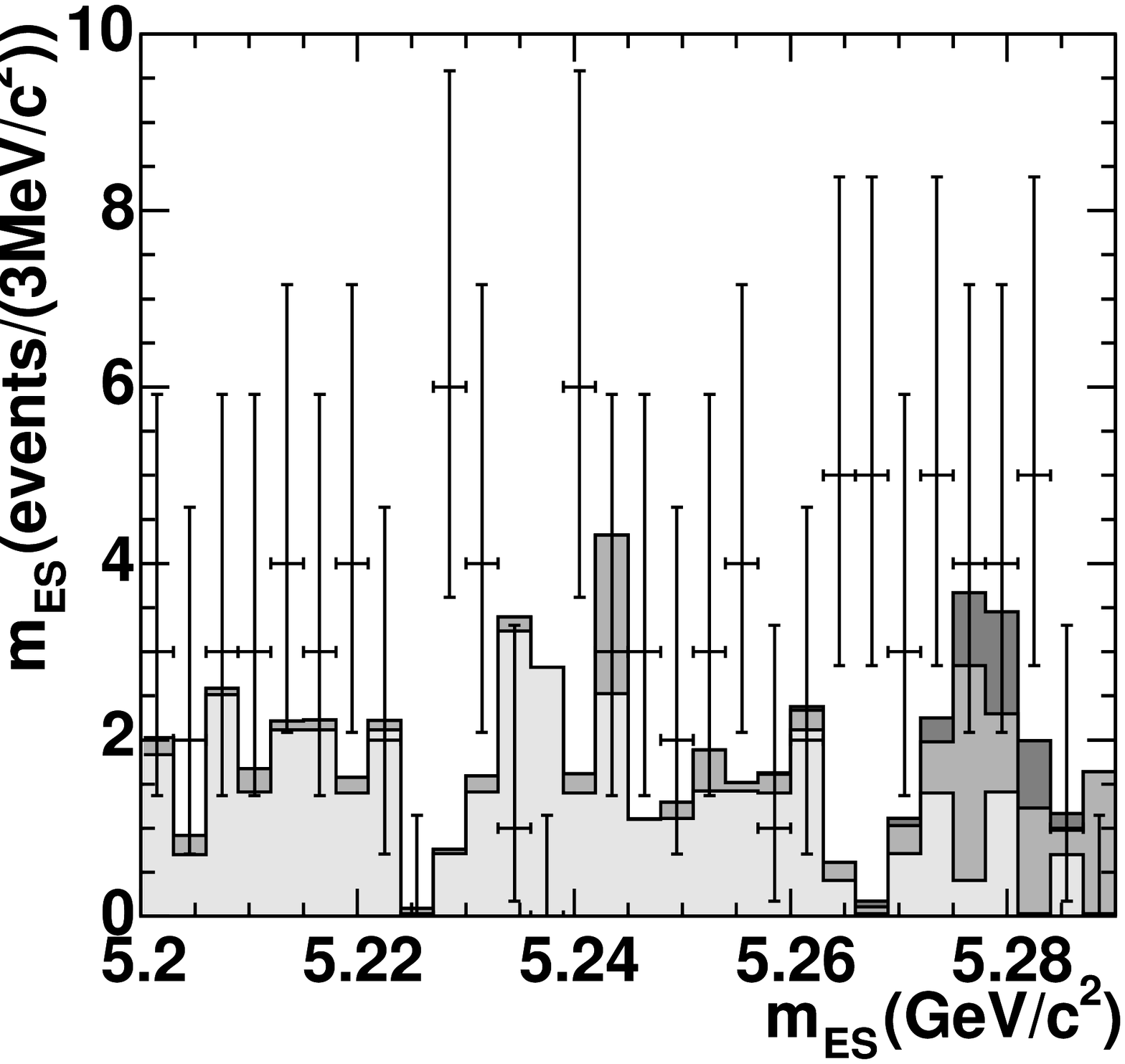}}
\vskip -.2 cm
\caption[]{\label{fig:1D}\small \DeltaE (left) and \mes (right)
distributions for data (points) and MC simulation (shaded histograms).
All selection requirements are applied including the \mes signal box
requirement for the left plot and the \DeltaE signal box requirement for
the right plot.}
\end{figure}

We~observe \Nseen events in the signal box.  Figure~\ref{fig:1D}
presents the \DeltaE and \mes distributions with all selection
requirements applied.  The Monte Carlo simulation is shown with separate
contributions from \pizbackground, other \BB, and \qqbar events.

\begin{figure}[!thb]
\centerline{\epsfxsize 3.0truein \epsfbox{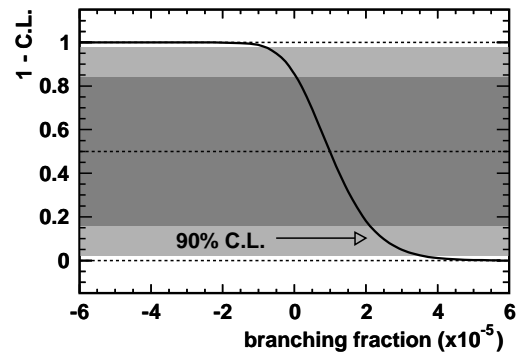}}
\vskip-4mm
\caption[]{\label{fig:prob}\small 1$-$confidence level versus the
assumed branching fraction.  The~shaded areas are the 68\% and 95\%
probability regions.  The~90\%~C.L. is marked with an arrow.}
\end{figure}

The~branching fractions are determined in a frequentist-model approach,
modified based on Ref.~\cite{bib:RogerBarlow}.  Besides taking the
systematic uncertainty in the efficiency and the statistical uncertainty
in the background estimate into account, the background expectation
value is also shifted by a factor selected from a flat distribution of
the range determined by the data-to-Monte Carlo ratios (see
Table~\ref{tab:results}).  When combining all six modes, this shift
comes from the range 0.4 to 1.3 (derived from $0.6\pm0.2$ and
$1.2\pm0.1$) and is applied coherently for each of the modes.  We~assume
that 50\% of the \FourS mesons decay into neutral \BB pairs.
Figure~\ref{fig:prob} displays 1$-$C.L. versus the assumed branching
fraction.  The~significance of this measurement, {\em i.e.}, 1$-$C.L. at
branching fraction zero, is \TheSignificance.  The~central value of the
branching fraction of \mydecay is \TheCV, which is consistent with zero.
The~upper limit on the branching fraction is $\BR(\mydecay)<\;\TheLimit$
at 90\% confidence level and is in agreement with the theoretical
expectations.

We are grateful for the excellent luminosity and machine conditions
provided by our \pep2\ colleagues, 
and for the substantial dedicated effort from
the computing organizations that support \babar.
The collaborating institutions wish to thank 
SLAC for its support and kind hospitality. 
This work is supported by
DOE
and NSF (USA),
NSERC (Canada),
IHEP (China),
CEA and
CNRS-IN2P3
(France),
BMBF and DFG
(Germany),
INFN (Italy),
FOM (The Netherlands),
NFR (Norway),
MIST (Russia), and
PPARC (United Kingdom). 
Individuals have received support from CONACyT (Mexico), A.~P.~Sloan Foundation, 
Research Corporation,
and Alexander von Humboldt Foundation.

\begin{thebibliography}{99}
\bibitem{bib:CC} Charge-conjugates are implied throughout this paper.
\end{thebibliography}
\end{document}